\documentclass[journal]{IEEEtran}

\makeatletter
\def\endthebibliography{%
	\def\@noitemerr{\@latex@warning{Empty `thebibliography' environment}}%
	\endlist
}
\makeatother

\usepackage{ifpdf}

\usepackage{cite}
\usepackage{subcaption}
\usepackage[abs]{overpic}

\usepackage{graphicx}
\usepackage{epstopdf} 
\ifCLASSINFOpdf
\else
\fi

\usepackage[cmex10]{amsmath}
\usepackage{amsfonts,amssymb}
\usepackage{xcolor}

\usepackage{algpseudocode}
\usepackage{algorithm}

\usepackage{array}

\usepackage{expl3}
\ExplSyntaxOn
\newcommand\latinabbrev[1]{
	\peek_meaning:NTF . {
		#1\@}%
	{ \peek_catcode:NTF a {
			#1.\@ }%
		{#1.\@}}}
\ExplSyntaxOff

\usepackage{dblfloatfix}

\usepackage{newtxtext, newtxmath}
\usepackage{url}
\usepackage{arydshln}
\usepackage{enumerate}
\usepackage[mathcal]{euscript}

\newtheorem{lemma}{Lemma}

\newtheorem{assumption}{\textbf{Assumption}}
\newtheorem{remark}{\textbf{Remark}}
\newcommand{\vect}[1]{\mathbf{#1}}

\def\beq{\begin{equation}}
\def\eeq{\end{equation}}
\def\vp{\varphi}
\def\etc{\latinabbrev{etc}}

\newcommand{\bs}[1]{\boldsymbol{#1}}

\DeclareMathOperator*{\argmin}{argmin}

\graphicspath {{Figures/}}
\IEEEoverridecommandlockouts

\begin{document}
%
\title{What Role Can NOMA Play in Massive MIMO?}
%
%
%

\author{\IEEEauthorblockN{Kamil Senel, Hei Victor Cheng, Emil Bj\"{o}rnson, and Erik G.~Larsson} 
	\thanks{The authors are with the Department of Electrical Engineering (ISY), Link\"{o}ping University, Sweden.}
	\thanks{This work was supported in part by ELLIIT, Swedish Research Council (VR).}
}

%
%


\maketitle


\begin{abstract}
This paper seeks to answer a simple but fundamental question: What role can NOMA play in massive MIMO? It is well-established that power-domain non-orthogonal multiple access (NOMA) schemes can outperform conventional orthogonal multiple access (OMA) schemes in cellular networks. However, this fact does not imply that NOMA is the most efficient way to communicate in massive MIMO setups, where the base stations have many more antennas than there are users in the cell. These setups are becoming the norm in future networks and are usually studied by assuming spatial multiplexing of the users using linear multi-user beamforming.
To answer the above question, we analyze and compare the performance achieved by NOMA and multi-user beamforming in both NLOS and LOS scenarios. We reveal that the latter scheme gives the highest \emph{average} sum rate in massive MIMO setups. We also identify specific cases where NOMA is the better choice in massive MIMO and explain how NOMA plays an essential role in creating a hybrid of NOMA and multi-user beamforming that is shown to perform better than two standalone schemes do.
\end{abstract}


%
\IEEEpeerreviewmaketitle

\section{Introduction}\label{sec:Introduction}

\IEEEPARstart{T}{he} fifth generation ($5$G) cellular technology aims to handle the $1000$-fold increase in mobile data traffic that is predicted over the next decade. A vital challenge for $5$G networks is to greatly improve the spectral efficiency compared to contemporary networks. One of the key technologies to increase the per-cell spectral efficiency is non-orthogonal multiple access (NOMA) which has attracted a considerable attention of researchers \cite{ding2017survey}. Another key technology is massive MIMO, which refers to the use of base stations (BSs) with a large number of antennas \cite{Larsson2014a}.
Now the first release of the $5$G standard has been finalized and contains a variety of features, including NOMA and massive MIMO. It is important to determine which features provide the largest performance gains in practical scenarios.
In this paper, we investigate the application of NOMA at BSs that are equipped with many antennas, since that appears to be the norm in the new LTE-Advanced and 5G deployments \cite{Liu2017a,Sprint2018a}.

The key idea of NOMA is to serve multiple users at the same time/frequency/code resource and thereby increase the sum spectral efficiency in the cell. The typical approach to NOMA is to group users and superpose their data signals using different transmission powers before transmitting the group's signal in the same way, using the same beamforming. The users are usually grouped to have very different channel conditions (e.g,. one cell-center user is grouped with one cell-edge user). Users with poor channel gains are allocated more transmission power to (partially) overcome their poor channel conditions and interference created by transmissions to other users. In each group, the user with the better channel can decode the signal sent to the user with the poorer channel, and the interference can thus be 
eliminated by a process called successive interference cancellation (SIC) \cite{ding2017application}. Hence, the user with the better channel does not need to be 
allocated a high transmission power to achieve a good communication rate.
This approach is sometimes referred to as power-domain NOMA.

There are two crucial drawbacks of SIC. The first and foremost drawback is that accurate channel state information (CSI) is required at the receiver side to utilize SIC, without being subject to substantial residual interference. Hence, downlink pilot signaling is  essential for NOMA and the performance of SIC relies heavily on the channel estimation quality \cite{usman2016performance}. 
Another drawback is the additional computational complexity and buffering of the received signals that are needed by the SIC procedure. The computational burden on the user devices may be too demanding for NOMA systems \cite{ding2017application}, at least for some applications. 

As indicated by the NOMA terminology, it constitutes a break with the traditional orthogonal multiple access (OMA) approaches (e.g., used in OFDMA and TDMA), where resources are exclusively allocated to users and hence, there is no intra-cell interference when using OMA. While convenient to implement, the OMA techniques suffer from inefficient utilization of the spectral resources. OMA is the common benchmark for the performance assessment of NOMA techniques and there are prior works showing that NOMA is always better than OMA in terms of sum rate, under the strong assumption of having perfect CSI. 
There are also a number of prior works that provide comparisons of NOMA and OMA  in more realistic setups with imperfect CSI \cite{yang2016performance,wei2017optimal}. 
Our goal is not to compare NOMA and OMA, but to investigate how NOMA can be utilized by BSs equipped with many antennas.

In less than a decade, massive MIMO has transitioned from being a far-fetched theoretical concept with an unlimited number of antennas \cite{Marzetta2010a} to a practical technology that has been commercially deployed in LTE-Advanced networks using 64-antenna BSs \cite{Sprint2018a}. In a nutshell, massive MIMO refers to systems where the BSs are equipped with a large number of antennas, $M$, as compared to the number of simultaneously active users, $K$. In other words, $M \gg K$ is the \emph{characterization of a massive MIMO setup}. The BS antennas are used for spatial multiplexing of the users at the same time/frequency/code resource \cite{Larsson2014a,massivemimobook}. Each user is assigned a dedicated beam that is adapted to the collection of user channels, in order to balance between achieving a high array gain for the desired signal and limiting the inter-user interference \cite{bjornson2014optimal}. Zero-forcing (ZF) is a popular interference-suppressing beamforming scheme in massive MIMO since it eliminates all the inter-user interference under perfect CSI and performs well in practical situations with  imperfect CSI, where substantial residual interference remains.
Hence, even though massive MIMO has sometimes been referred to as {spatial} OMA \cite{chen2016application}, massive MIMO is definitely a non-orthogonal multiple access technology. However, for clarity, the NOMA abbreviation will be exclusively used to refer to the power-domain NOMA scheme throughout this paper.   

While the vast majority of prior works on NOMA considers single-antenna BSs, there are some papers that consider $M$-antenna BSs. 
NOMA and OMA are compared with a small $M$ in \cite{chen2017exploiting, ding2017application,sun2015ergodic} and a relatively large $M$ in \cite{cheng2018performance,ding2016design}. However, in these scenarios, OMA and (power-domain) NOMA  techniques are not the only multiple access techniques to be considered, but traditional multi-user MIMO beamforming (e.g., based on ZF) must also be included  to figure out the most spectrally efficient way to communicate.
Such a comparison can be found in \cite{chen2016application,chen2016beamforming} for $M \approx K$, which is not a massive MIMO setup, and in \cite{cheng2018performance} for $M \gg K$ but using the vastly suboptimal maximum-ratio processing scheme.
Hence, to the best of our knowledge, the performance of (power-domain) NOMA has not been properly analyzed in massive MIMO setups with $M \gg K$ and compared with state-of-the-art massive MIMO methods.

\subsection{Main Contributions}

Bearing in mind that many pre-5G and 5G deployments are considering massive MIMO setups with $M \gg K$ \cite{Liu2017a,Sprint2018a}, in this work we consider a single-cell system with $M \gg K$ and compare the performance of a typical NOMA scheme with a typical ZF-based massive MIMO beamforming scheme, which we abbreviate as the mMIMO scheme.
 Both non-line-of-sight (NLOS) and line-of-sight (LOS) channel models are considered. We identify the situations in which the different schemes prevail and, in the final part of the paper, we consider a combination of the two schemes that exploits these insights.
 To summarize, the main contributions of this work are as follows:
\begin{itemize}
	\item We provide a performance analysis and comparison of the NOMA and mMIMO schemes in a massive MIMO setup with NLOS channels, showing that mMIMO outperforms NOMA in terms of sum rate.
	\item Closed-form expressions for the maximum sum rate for a two user setup are provided to achieve insights into the performance of NOMA and mMIMO in NLOS. Moreover, the minimum number of BS antennas required for mMIMO scheme to outperform NOMA is derived.
	\item The analysis is extended to LOS channels. By solving a sum rate optimization problem we see that mMIMO significantly outperforms NOMA in terms of average sum rate.  
	\item We demonstrate that even though mMIMO is better than NOMA when averaging the sum rate over different user locations, there is a non-negligible probability that NOMA performs better for a particular collection of users. 
	\item We show that by employing a hybrid mMIMO-NOMA scheme, it is possible to obtain a better overall performance compared to the standalone mMIMO and NOMA scheme.
\end{itemize}

\section{System Setup}\label{sec:SystemSetup}
 
We consider the downlink transmission in a single-cell system with an $M$-antenna base station (BS) and $K$ single-antenna users. The user set, $\mathcal{K}$, consists of $K/2$ cell-edge users and $K/2$ cell-center users. Here, we assume $K$ is even and the indexes $k \in \mathcal{K}_c$, where $\mathcal{K}_c = \{1, \ldots, K/2 \}$, are utilized for cell-center users and $k \in \mathcal{K}_e = \{K/2+1, \ldots, K\}$ denotes cell-edge users. This classification of cell-edge and cell-center users is not strict, but we are merely dividing the users into two sets, such that \eqref{eq:largeScaleAssumption} below holds.

The channel vector of user $k$, $\vect{g}_k$ is modeled as
\begin{equation}\label{eq:NLOSchannel}
\vect{g}_k = \sqrt{\beta_k}\vect{h}_k, \quad k = 1, \ldots,K, 
\end{equation}
where $\beta_k$ represents the large-scale fading coefficient which is assumed to be known at the BS, and satisfy 
\begin{equation}\label{eq:largeScaleAssumption}
\beta_j > \beta_i, \quad \forall j \in \mathcal{K}_c,~~ \forall{i} \in \mathcal{K}_e.
\end{equation}
Notice that the grouping of users into cell-center and edge users may not necessarily reflect their distances to the BS in a real system, i.e., a user far from the BS may have a higher large-scale fading coefficient than a user close to the BS due to shadowing and may belong to $\mathcal{K}_c$.

\begin{remark}
 Notice that, the effects of shadow fading can be incorporated in the large-scale fading coefficients. However, it is not explicitly considered in the numerical analysis.
\end{remark}

We consider both  line-of-sight (LOS) and non-line-of-sight (NLOS) communication. 
In the LOS case, $\vect{h}_k$ is an arbitrary constant/deterministic vector, which is perfectly known at the BS since deterministic variables can be estimated with a negligible estimation overhead.

In the NLOS case, we consider a block fading model where the time-frequency resources are divided into coherence intervals in which the channels are constant and frequency flat. 
We let $T$ denote the total number of symbols (i.e., channel uses) per coherence interval.
The system operates in time-division duplex (TDD) mode so that channel reciprocity can be utilized at the BS to estimate the downlink channels based on uplink pilots, and the BS later uses these estimates for downlink multiuser beamforming.
In each coherence interval, $\vect{h}_k$ in \eqref{eq:NLOSchannel} for user $k$ takes one independent random small-scale fading realization from an independent Rayleigh fading distribution, i.e., 
\begin{equation}\label{eq:RayleighFading}
\vect{h}_k \sim \mathsf{CN}\,(\vect{0},\vect{I}_M), \quad k = 1, \ldots,K.
\end{equation} 
This is the realization that the BS wishes to estimate, and we will deal with the estimation in detail in Section~\ref{sec:ChEstimation}.

\subsection{NOMA and mMIMO Schemes} \label{subsec:schemes-definition}

In this paper, we will compare the achievable performance achieved by one typical NOMA scheme and one typical mMIMO scheme. We consider a \emph{NOMA scheme} that has been analyzed in various works in the literature \cite{cheng2018performance,kim2013non}. The $K$ users are grouped into $K/2$ groups, where each group consists of two users: one cell-center and one cell-edge user. Without loss of generality, assume that the indices $i$ and $i + K/2$ denote the users in group $i$ for $i = 1, \ldots, K/2$.

In the NOMA scheme, the interference between groups can be eliminated by either allocating orthogonal resources to different groups or using judiciously selected beamforming vectors, which is assumed in this work. 
Let $\vect{v}_i$ denote the normalized beamforming vector for group $i$, which is generated based on the cell-center users' channels. The beamforming vectors are selected to cancel interference between the cell-center users, resulting in
\begin{equation}\label{eq:conditionBF}
\vect{v}_j^H\vect{h}_i = \begin{cases}
~0, &\text{if}~j\neq i, \\
~1, &\text{if}~j = i,
\end{cases}~~\forall j = 1,\ldots,K/2,
\end{equation}
for all $i$ \cite{chen2017exploiting,kim2013non}. The rationale for this choice is that the cell-center user is sensitive to interference from other groups since it needs to be able to perform SIC to cancel interference from the cell-edge user in the own group.\footnote{There are two users and one beamforming vector to be generated for each group. It is possible to generate the beamforming vector based on a linear combination of the channels in a group \cite{cheng2018performance}. However, such an approach requires estimation of a linear combination users' channels. Especially, for the cases, such as mMTC, where one user aims for low rates, spending system resources to acquire their channels may not be efficient.} To implement this type of ZF beamforming, the BS only needs to know the $K/2$ channels to the cell-center users.


\begin{figure} 
	\begin{tabular}{cc}
		\begin{subfigure}{.2\textwidth}
		\includegraphics[width=40mm]{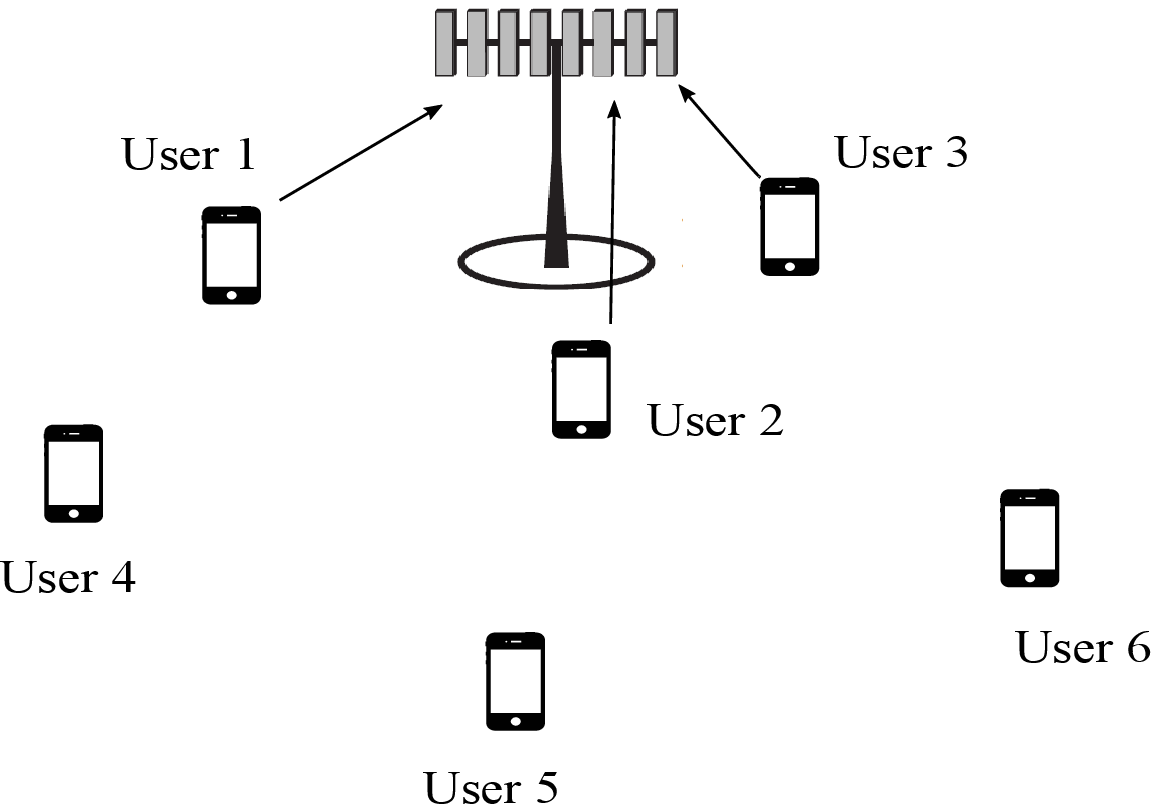}
		\caption{} \label{fig:NOMAtraining}
	\end{subfigure} & \hfill \begin{subfigure}{.2\textwidth}\begin{center} 
	\includegraphics[width=40mm]{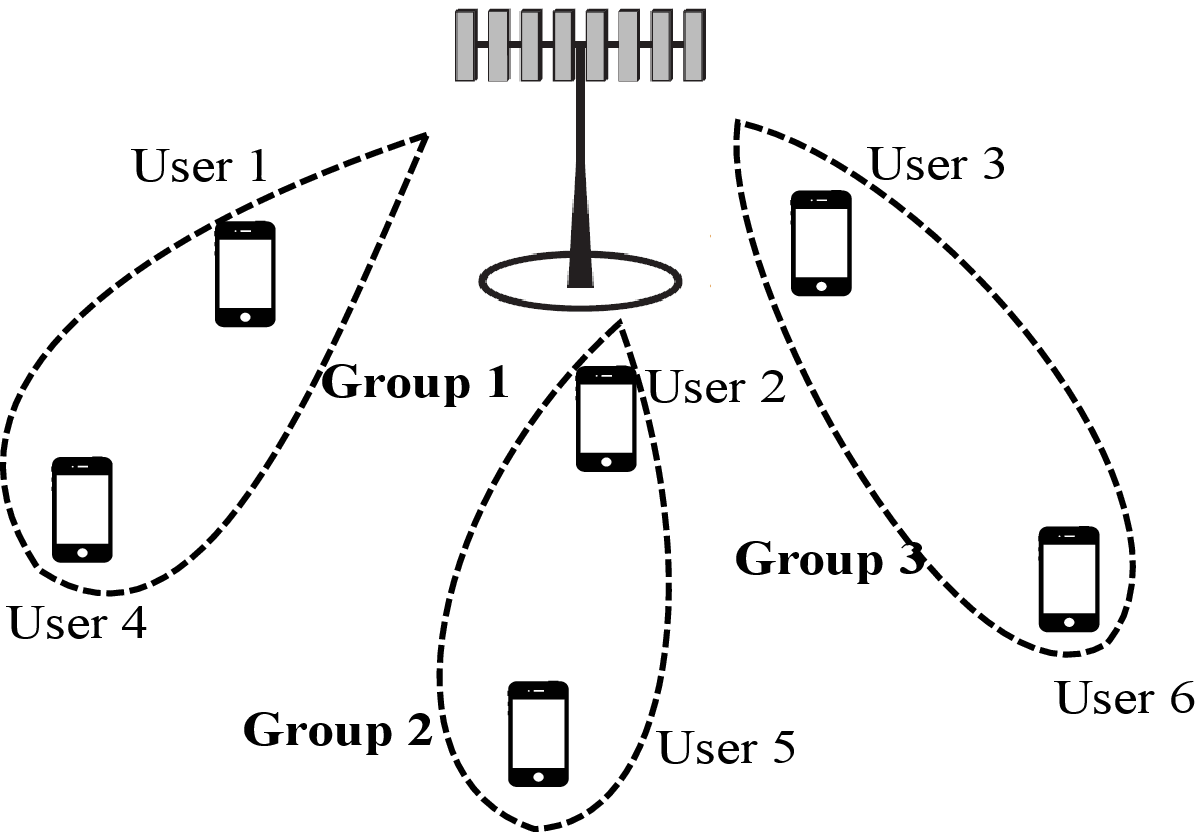}
	\caption{} \label{fig:NOMAdata} \end{center}
\end{subfigure} \\ \vspace{-1mm} & \vspace{-1mm}\\
		\begin{subfigure}{.2\textwidth}
			\includegraphics[width=40mm]{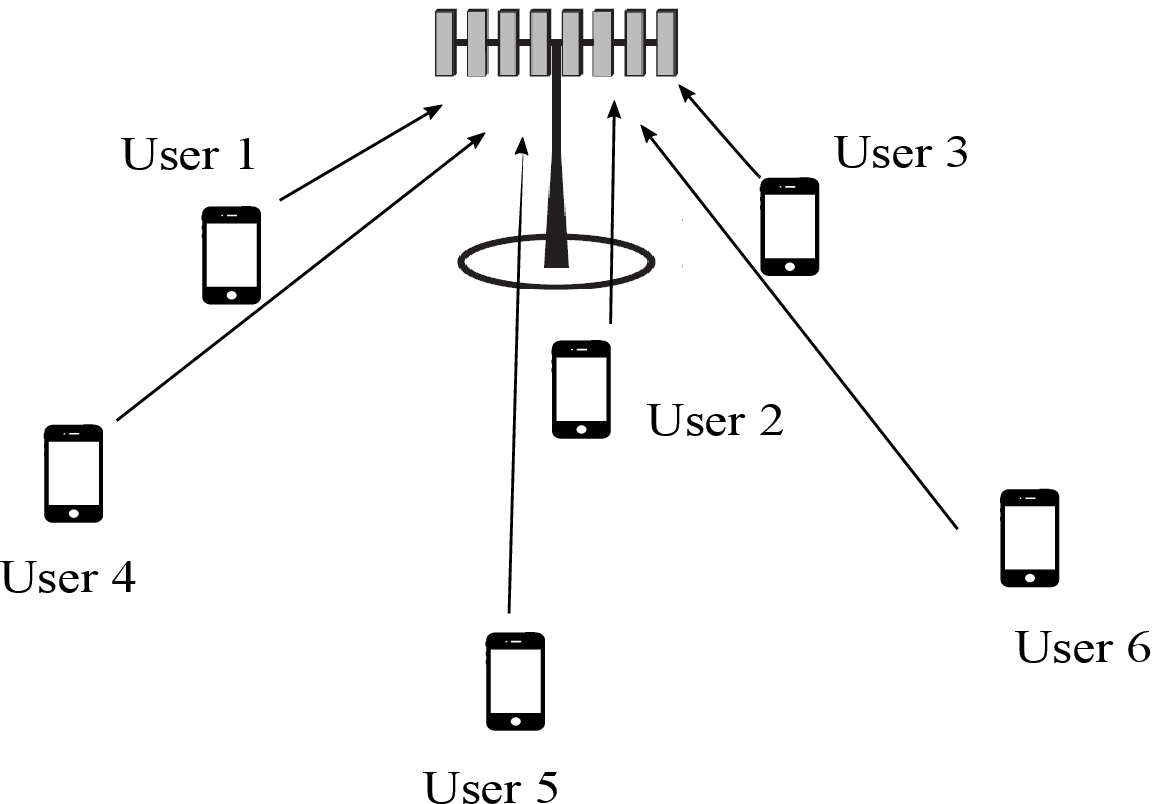}
			\caption{} \label{fig:ZFtraining}
		\end{subfigure} & \hfill \begin{subfigure}{.2\textwidth}
		\includegraphics[width=40mm]{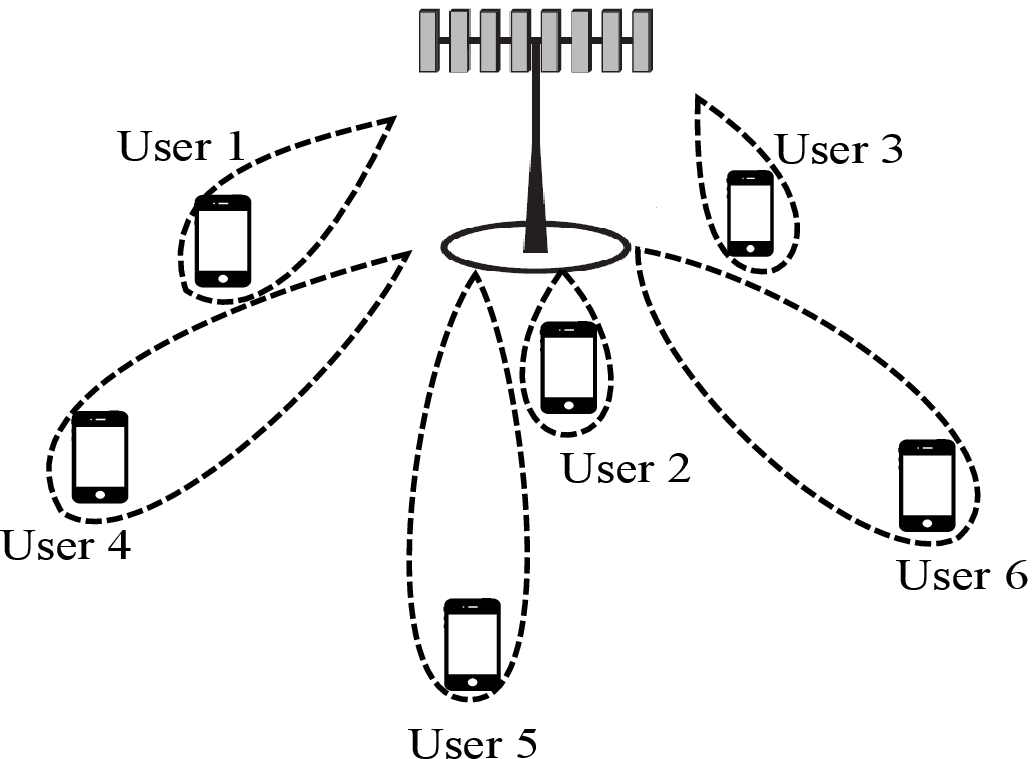}
		\caption{} \label{fig:ZFdata}
	\end{subfigure} 
	\end{tabular}
	\caption{The training and data transmission phases for mMIMO and NOMA with $K = 6$ users. (a) During the training of NOMA cell-center users transmit pilot sequences, while the cell-edge users are silent, and the BS estimates their channels. (b) Users in the same NOMA group receives data via the same beamforming vector. (c) In mMIMO, each user transmits its pilot sequence during training and the BS estimates the channels for every user. (d) Each user has a separate mMIMO beamforming vector generated by using the estimate of their channels.   }
	\label{fig:NOMA-ZFsetups}
\end{figure}

The other considered scheme is called the \emph{mMIMO scheme} and is based on ZF beamforming, which is a common assumption in the mMIMO literature and nearly optimal in single-cell systems with many antennas \cite{bjornson2014optimal}. In the mMIMO case, unlike NOMA, there are no groups. There are $K$ beamforming vectors, where $\vect{v}_k$ is the beamforming associated with user $k$, and these are selected based  on $K$ channels instead of $K/2$ channels, as in NOMA. Hence, the BS needs to know $K$ channels. In the mMIMO scheme, the ZF beamforming vectors satisfy
\begin{equation}\label{eq:conditionBFZF}
\vect{v}_j^H\vect{h}_i = \begin{cases}
~0, &\text{if}~j\neq i, \\
~1, &\text{if}~j = i,
\end{cases}~~\forall j = 1,\ldots,K,
\end{equation}
for all $i$.

\begin{remark} \label{remark-NOMA}
Throughout the paper, we will be making a series of assumptions that favor NOMA as our goal is to find conditions where NOMA might perform better than the mMIMO scheme. With this methodology, we can be sure that mMIMO provides higher rates than NOMA whenever the analytical results show that. In the massive MIMO region ($M \gg K$), mMIMO scheme significantly outperforms NOMA on average, however, as will be demonstrated later, mMIMO systems can still benefit by employing NOMA in some cases.
\end{remark}

\subsection{Pilot Overhead for Estimating NLOS Channels}\label{sec:ChEstimation}

In the NLOS scenario, the channels need to be estimated frequently and therefore the channel acquisition overhead cannot be neglected (as is the case in the LOS scenario).
In a traditional TDD system, each coherence interval consists of three phases: uplink training, uplink data transmission, and downlink data transmission. In this work, our focus is on downlink data transmission, and uplink data transmission is not considered.
 During the uplink training, some or all of the users transmit pilot sequence and the BS uses them to estimate the channels. Then, these estimates are used to generate the beamforming during data transmission. 

The considered NOMA scheme requires $K/2$-length pilot sequences in the uplink, giving room for one orthogonal pilot per cell-center user. The training and data transmission phases are illustrated in Fig.~\ref{fig:NOMA-ZFsetups}(a)-(b) in an example with $K = 6$ users. To perform SIC, the cell-center users need to learn the effective channels that are created by the beamforming, otherwise NOMA provides no advantage over conventional OMA approaches \cite{cheng2018performance}. Moreover, since the beamforming vectors are only based on the cell-center users' channels, the effective channels of the cell-edge users will fluctuate substantially between coherence intervals and the phase will be uniformly distributed from $-\pi$ to $+\pi$. Hence, the BS needs to send $K/2$ pilot sequences in the downlink to let the receiving users estimate their effective channels when using the NOMA scheme. We also make the following assumption.

\begin{assumption}\label{as:DLpilots}
In the analysis of the NOMA scheme, it is assumed that perfect CSI is acquired at the users via downlink pilots. 	
\end{assumption}

In contrast, the mMIMO scheme requires $K$-length pilot sequences in the uplink, giving room for one orthogonal pilot per user. The training and data transmission phases are illustrated in Fig.~\ref{fig:NOMA-ZFsetups}(c)-(d) in an example with $K = 6$ users. 
An important advantage of mMIMO  is that downlink pilots are not needed since the effective channels that are created by the beamforming are highly predictable (nearly deterministic gain and phase due to channel hardening), as further explained in \cite{ngo2017no, caire2018ergodic}. Hence, no downlink pilots are needed for the mMIMO scheme.
\begin{figure}
\begin{overpic}[trim=0.5cm 0cm 0cm 0cm,clip=false, scale = .5]{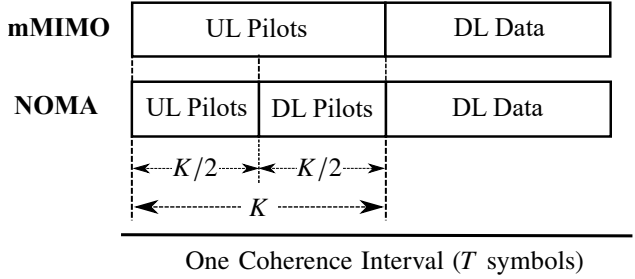}
	\put(55,37){$K/2$} \put(102,37){$K/2$}
	\put(81,22){\colorbox{white}{$K$}}
	\put(57,2){\colorbox{white}{One Coherence Interval ($T$ symbols)}}
\end{overpic}
\caption{Coherence interval structure for the considered mMIMO system. Upper and bottom figures illustrate the training and data transmission structure for the mMIMO and NOMA schemes under NLOS setup.}\label{fig:CIs} 
\end{figure}


In summary, $K$ pilots are needed for both schemes, but they are allocated differently between uplink and downlink; see Fig.~\ref{fig:CIs} for an illustration. 
We will now provide the mathematical details of the uplink training, which will later be important when quantifying the achievable rates.

\subsubsection{Channel Estimation with mMIMO}

In this case, all the users transmit pilots. Let $\sqrt{K}\bs{\vp}_k \in \mathbb{C}^{K \times 1}$ denote the $K$-length orthogonal pilot sequence of user $k$, and
for all $k \in \mathcal{K}$:
\begin{equation}
\bs{\vp}_k^H\bs{\vp}_j = \begin{cases}
1, &\text{if}~j = k, \\
0, & \text{if}~j \neq k, 
\end{cases}~~ \forall j \in \mathcal{K}.   
\end{equation}
Note that the total pilot energy expended by each user scales with the length of the pilot sequences. The received signal at the BS is given by
\begin{equation}\label{eq:recSigZF}
\vect{Y} = \sqrt{K}\sum\limits_{k' \in \mathcal{K}} \sqrt{q_{k'}}\vect{g}_{k'} \bs{\varphi}_{k'}^H + \vect{Z}.
\end{equation}
 Here, the normalized transmission power of the pilot symbols for user $k$ is denoted by $q_{k}$ and $\vect{Z} \in \mathbb{C}^{M \times K}$ is the noise matrix with i.i.d.$~\mathsf{CN}\,(0,\,1)$ elements. 

To acquire an estimate of the channel of user $k$, the BS performs a de-spreading operation as follows:
\begin{eqnarray}
\vect{y}_k &=& \vect{Y}\bs{\varphi}_k, \nonumber \\ 
&=& \sqrt{K q_k} \vect{g}_k + \sqrt{K}\hspace{-0.4cm}\sum\limits_{k' \in \mathcal{K}\backslash\{k\}} \hspace{-0.4cm}\sqrt{q_k}\vect{g}_{k'} \bs{\varphi}_{k'}^H\bs{\varphi}_k  + \vect{z}', \nonumber \\
&=&\hspace{-2mm} \sqrt{K \beta_kq_k} \vect{h}_k + \vect{z}' , \label{eq:despred-1}
\end{eqnarray} 
where $\vect{z}' = \tilde{\vect{Z}}\bs{\varphi}_k$ has i.i.d.$~\mathsf{CN}(0,\,1)$ elements, since $\|\bs{\varphi}_k\|^2 = 1$. The de-spreading operation provides a noisy version of the channel and the BS utilizes the minimum mean-square error (MMSE) estimator to obtain the channel estimate $\hat{\vect{h}}_k^{\textrm{mMIMO}}$ with mMIMO as
\begin{equation}\label{eq:MMSEestZF}
\hat{\vect{h}}_k^{\textrm{mMIMO}} = \frac{\sqrt{K \beta_kq_k}}{K\beta_{k}q_{k} + 1}\vect{y}_k. 
\end{equation}  
This estimate has $M$ i.i.d.  $\mathsf{CN}(0, \gamma_{k}^{\textrm{mMIMO}})$ elements with variance
\begin{align}\label{eq:MSgammZF}
\gamma_{k}^{\textrm{mMIMO}} =  \frac{K\beta_kq_k}{K\beta_{k}q_{k} + 1}.
\end{align} 

\subsubsection{Channel Estimation with NOMA}

During uplink training each cell-center user transmits $K/2$-length orthogonal pilots and the cell-edge users remain silent. For NOMA, the received signal at the BS during uplink training is 
\begin{equation}\label{eq:recSigNOMA}
\vect{Y} = \sqrt{K/2}\sum\limits_{k' \in \mathcal{K}_c} \sqrt{q_{k'}}\vect{g}_{k'} \bs{\varphi}_{k'}^H + \vect{Z},
\end{equation}
where $\bs{\varphi}_{1},\ldots,\bs{\varphi}_{K/2} \in \mathbb{C}^{K/2}$ are the orthogonal pilot sequences.

  
Similarly to the mMIMO case, the MMSE channel estimate of cell-center user $k$ is
 \begin{equation}\label{eq:MMSEestNOMA}
 \hat{\vect{h}}_k^{\textrm{NOMA}} = \frac{\sqrt{K \beta_kq_k/2}}{K\beta_{k}q_{k}/2 + 1}\vect{Y}\bs{\varphi}_k,
 \end{equation}  
 where $\vect{Y}$ is given in \eqref{eq:recSigNOMA} when using NOMA. The estimate 
 $ \hat{\vect{h}}_k^{\textrm{NOMA}}$  has i.i.d. $\mathsf{CN}(0, \gamma_{k}^{\textrm{NOMA}})$ elements with  
 \begin{align}\label{eq:MSgammNOMA}
 \gamma_{k}^{\textrm{NOMA}} = \frac{K\beta_kq_k}{K\beta_{k}q_{k} + 2}. 
 \end{align}


\section{Performance Analysis for NLOS} \label{sec:NLOS}

In this section, we analyze and compare the achievable rates of the NOMA and mMIMO schemes when using the NLOS channel model. The channel estimates described in Section~\ref{sec:ChEstimation} are utilized to generate the downlink beamforming vectors. 


\subsection{Downlink Data Transmission}

After the training phase, the BS generates beamforming vectors, based on the ZF criteria described in Section~\ref{subsec:schemes-definition},
and user $k$ receives
\begin{equation}\label{eq:RecSignal}
y_k = \sum_{k' = 1}^{K} \vect{g}_k^T\vect{x}_{k'} + z_k,
\end{equation}
where $z_k \sim \mathsf{CN}\,(0,\,1)$ is the additive noise and $\vect{x}_k \in \mathbb{C}^M$ is the beamformed data symbol of user $k$ obtained as
\begin{equation}
\vect{x}_k = \vect{v}_k \sqrt{p_k}s_k.
\end{equation}
Here, $s_k\sim\mathsf{CN}(0, 1)$ is the data symbol of user $k$; $p_k$ is the normalized transmission power of data symbols for user $k$. This modeling applies to both the NOMA and mMIMO schemes, but the beamforming vectors are selected differently.
In the NOMA scheme, the beamforming vector of the users in the same group are identical, i.e., $\vect{v}_k = \vect{v}_{k + K/2}$ for all $k = 1, \ldots, K/2$. In contrast, each user has a unique beamforming vector in the mMIMO scheme. 

\subsubsection{Rates with NOMA}

The received signal in \eqref{eq:RecSignal} can be written as 
\begin{eqnarray}\label{eq:RecSignal-2}
y_k &=& \sqrt{\beta_{k}}\vect{h}_k^T\sum_{k' = 1}^{K} \vect{v}_{k'} \sqrt{p_{k'}}s_{k'} + z_k, \\
    &=& \sqrt{\beta_{k}p_k}\vect{h}_k^T \vect{v}_{k} s_{k} + \sqrt{\beta_{k}}\vect{h}_k^T\sum\limits_{\substack{k' \neq k,\\ k' = 1}}^{K} \vect{v}_{k'} \sqrt{p_{k'}}s_{k'} + z_k.
    \label{eq:RecSignalGeneral}
\end{eqnarray}
and the instantaneous SINR of users in group $k$ is 
\begin{equation} \label{eq:SINRNOMAnoSIC}
\text{SINR}_k = \frac{p_k\beta_k|\vect{h}_k^T \vect{v}_{k}|^2}{\beta_k \sum_{k' \neq k}^{K} p_{k'}|\vect{h}_k^T \vect{v}_{k'}|^2 + 1}
\end{equation} 
under Assumption~\ref{as:DLpilots}.

In each group, the cell-edge users treat the interference as noise and decodes its own data symbols, whereas the cell-center user can decode the data symbols of the cell-edge user and perform SIC, hence effectively removing the interference due to the cell-edge user. However, in order to perform SIC, the cell-center user needs to be able to decode data signal intended for the cell-edge user, i.e., ergodic achievable rate of the data signal of the cell-edge user, $s_{k+K/2}$, at user $k$ must be greater than or equal to the ergodic achievable rate of the cell-edge user. That is, the following condition must be satisfied: 
\begin{equation}\label{eq:SICcond}
\mathbb{E}\left[\log_2 \left(1 + \text{SINR}_{k,k + K/2} \right)\right]  \geq \mathbb{E}\left[\log_2 \left(1 + \text{SINR}_{k + K/2} \right)\right],
\end{equation} 
where 
\begin{equation}
\text{SINR}_{k,k + K/2} =  \frac{p_{k+K/2}\beta_{k}|\vect{h}_k^T \vect{v}_{k + K/2}|^2}{\beta_k \sum_{k' \neq k + K/2}^{K} p_{k'}|\vect{h}_k^T \vect{v}_{k'}|^2 + 1}.
\end{equation}
This condition can always be satisfied by selecting the transmit powers appropriately, thus we make the following assumption.

\begin{assumption}\label{as:SICcond}
	In the NOMA scheme, the SIC condition, defined by \eqref{eq:SICcond}, is assumed to be satisfied for each group. 
\end{assumption}

Under Assumptions \ref{as:DLpilots} and \ref{as:SICcond}, the achievable ergodic rate of user $k$ is given by  
\begin{flalign}
R_{k}^{\mathrm{NOMA}}=\tau  \mathbb{E}\left[\log_2\left(1+\frac{p_k
	\beta_k |\vect{h}_k^T\vect{v}_k|^2}{\beta_k\hspace{-3mm}\sum\limits_{\substack{k'\neq k,\\ k' \neq k+K/2}}^K \hspace{-3mm} p_{k'}|\vect{h}_k^T\vect{v}_{k'}|^2+ 1}\right)\right], \forall k \in \mathcal{K}_c,\label{eq:rateNOMAcc}
\end{flalign}
where $\tau = \left(1-\frac{K}{T}\right)$ is the fraction of each coherence interval that is used for data.

For the cell-edge users, the achievable rate is given by 
\begin{equation}\label{eq:rateNOMAce}
R_{k}^{\mathrm{NOMA}}=\tau\mathbb{E}\left[\log_2 \left(1 + \text{SINR}_{k + K/2} \right)\right], \forall k \in \mathcal{K}_e,
\end{equation} 
where $\text{SINR}_{k + K/2}$ is defined in \eqref{eq:SINRNOMAnoSIC}.

A detailed derivation of the expressions in \eqref{eq:rateNOMAcc} and \eqref{eq:rateNOMAce} can be found in \cite{cheng2018performance}.

\subsubsection{Rates with mMIMO}

The mMIMO scheme has been throughly investigated in the literature  with using ZF in NLOS scenarios \cite{redbook, yang2013performance}. An ergodic achievable rate for user $k$ is given by
\begin{equation}\label{eq:rateZF}
R_{k}^{\mathrm{mMIMO}} \geq \tau\log_2\left(1 + \frac{\left(M-K\right)p_k\beta_{k}\gamma_{k}^{\mathrm{mMIMO}}}{ \beta_{k}(1 - \gamma_k^{\mathrm{mMIMO}})\sum\limits_{k'= 1}^K p_{k'}+1}\right),
\end{equation}
with ZF.

\subsection{Performance Comparison}

In this section, we compare the achievable rates given by \eqref{eq:rateNOMAcc}, \eqref{eq:rateNOMAce} and \eqref{eq:rateZF} with respect to the number of BS antennas, $M$ and the number of users $K$. While numerical analysis can be carried out based on the rate expressions, \eqref{eq:rateNOMAcc} and \eqref{eq:rateNOMAce}, in their current form, do not allow analytical comparison. Therefore, we first make the following assumption (recall Remark~\ref{remark-NOMA}).

\begin{assumption}\label{as:InterferenceGroups}
	For the NOMA scheme, the beamforming vectors are assumed to perfectly eliminate the interference between different groups. 
\end{assumption}

Using Assumption \ref{as:InterferenceGroups}, we obtain the following upper bounds for $k = 1, \ldots, K/2$, 
\begin{flalign}
&R_{k}^{\mathrm{NOMA}}\leq\tau  \mathbb{E}\left[\log_2\left(1+p_k
	\beta_k |\vect{h}_k^T\vect{v}_k|^2\sigma^2\right)\right], \label{eq:rateNOMAccUpperBound}
\end{flalign}
and 
\begin{flalign}
&R_{k}^{\mathrm{NOMA}}\leq\tau \mathbb{E}\left[\log_2\left(1+\frac{p_k
	\beta_k |\vect{h}_k^T\vect{v}_k|^2}{\beta_k p_{k-K/2}|\vect{h}_k^T\vect{v}_k|^2 + 1}\right)\right], \label{eq:rateNOMAceUpperBound}
\end{flalign}
for all $k = K/2 + 1, \ldots,K$. Next, we can use the Jensen's inequality on \eqref{eq:rateNOMAccUpperBound} and \eqref{eq:rateNOMAceUpperBound} as follows: 
\begin{flalign}
&R_{k}^{\mathrm{NOMA}}\leq\tau  \log_2\left(1+p_k
	\beta_k \mathbb{E}\left[|\vect{h}_k^T\vect{v}_k|^2\right]\right), \forall k \in \mathcal{K}_c \label{eq:rateNOMAccUpperBound-2}
\end{flalign}
and 
\begin{flalign}
&R_{k}^{\mathrm{NOMA}}\leq\tau  \log_2\left(1+\frac{p_k
	\beta_k \mathbb{E}\left[|\vect{h}_k^T\vect{v}_k|^2\right]}{\beta_k p_{k-K/2}\mathbb{E}\left[|\vect{h}_k^T\vect{v}_k|^2\right] + 1}\right), \forall k \in \mathcal{K}_e. \label{eq:rateNOMAceUpperBound-2}
\end{flalign}

\begin{remark}
	Note that we have started from achievable rate expressions that were developed using the same methodology, and then we derive a lower bound in \eqref{eq:rateZF} for mMIMO and upper bounds in \eqref{eq:rateNOMAccUpperBound-2} and \eqref{eq:rateNOMAceUpperBound-2}. This is in line with the methodology for comparing the two schemes that was explained in Remark~\ref{remark-NOMA}.
\end{remark} 

\subsection{Case study: Perfect CSI}

We first investigate the case where both uplink and downlink training results in perfect CSI. In this case, $\gamma_{k}^{\mathrm{mMIMO}} = 1$ and \eqref{eq:rateZF} becomes
\begin{equation}\label{eq:rateZFperfectCSI}
R_{k}^{\mathrm{mMIMO}} \geq \tau\log_2\left(1 + \left(M-K\right)p_k\beta_{k}\right),
\end{equation}
whereas \eqref{eq:rateNOMAccUpperBound-2} and \eqref{eq:rateNOMAceUpperBound-2} will respectively become
\begin{flalign}
&R_{k}^{\mathrm{NOMA}}\leq\tau  \log_2\left(1+p_k	\beta_k \left(M+1-K/2\right)\right), \forall k \in \mathcal{K}_c, \label{eq:rateNOMAccUpperBound-3}
\end{flalign}
and
\begin{flalign}
&R_{k}^{\mathrm{NOMA}}\leq\tau  \log_2\left(1+\frac{p_k
	\beta_k }{p_{k-K/2}\beta_k + 1}\right), \forall k \in \mathcal{K}_e.\label{eq:rateNOMAceUpperBound-3}
\end{flalign}

Notice the similarities between the rate expressions with the mMIMO and NOMA schemes, in terms of eliminating interference. In the case of perfect CSI, the mMIMO scheme eliminates interference completely at each user. On the other hand, NOMA eliminates interference at only one of the users in each group, namely the cell-center user. Moreover, all the $K$ users have a rate that grows with $M$ (known as the array gain) in the mMIMO scheme, while that only happens for the cell-center users in the NOMA scheme.

	\begin{figure}[tb]
	\begin{center}
		\includegraphics[trim=0.5cm 0cm 0cm 0cm,clip=false, scale = .7]{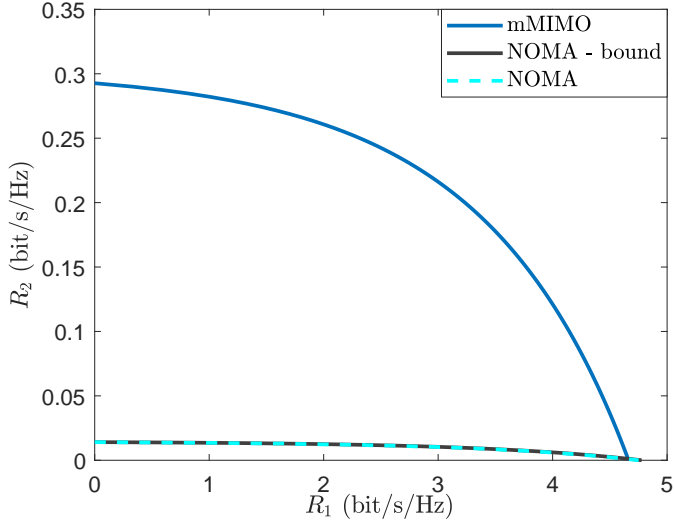}
		\caption{Rate regions obtained for two users with $M = 25$ BS antennas. The curves are obtained for a coherence interval length, $T = 100$ symbols under the perfect CSI assumption. }
		\label{fig:Fig1} 
	\end{center}
\end{figure}

Fig.~\ref{fig:Fig1} illustrates the rate regions for the mMIMO and NOMA schemes for $K = 2$ users and $M = 25$ BS antennas. The distance between the cell-center user and the BS is $100\,$m, whereas the cell-edge user is located at $350\,$m from the BS which results in approximately $20\,$dB received SNR difference at the users. In the figure, rate regions obtained using the closed-form rate bounds given in \eqref{eq:rateZF}, \eqref{eq:rateNOMAccUpperBound-3}, and \eqref{eq:rateNOMAceUpperBound-3}, are utilized and for NOMA they are compared with the actual rates defined by \eqref{eq:rateNOMAcc}, and \eqref{eq:rateNOMAce}. In the rest of the analysis, the lower bound on the rate is utilized for mMIMO, whereas the upper bounds are used for NOMA scheme. Notice that there is a significant difference between the rates achieved by the mMIMO and NOMA schemes for the cell-edge users. The reason that mMIMO outperforms NOMA at the cell-edge is that mMIMO provides the array gain to both users, while the NOMA scheme only give it to the cell-center user.     

 Next, we compare the performance of the mMIMO and NOMA schemes in terms of sum rate. Consider the sum rate of a group under NOMA scheme and assume for simplicity that indexes $1,2$ represent the cell-center and the cell-edge users, respectively. Then, the sum rate for the two users is  
\begin{eqnarray}
R_{\text{sum}}^{\mathrm{NOMA}} = \tau\log_2\left(1 + \bar{M}p_1\beta_1 \right) + \tau\log_2 \left(1 + \frac{p_2\beta_2}{1 + p_1\beta_2}\right),
\end{eqnarray}  
where  $\bar{M} = M + 1 - K/2$. The maximization of this sum rate is investigated in \cite{choi2016power} for two users.

Next, we generalize the two-user setup to $K$ users. Let $\vect{p} = [p_1,\ldots,p_K]^T$ denote the power vector. Then, the sum rate optimization problem can be stated as
      \begin{equation} \tag{P1}\label{pr:PR-1}
\begin{aligned}
& \underset{\vect{p}\succeq 0}{\text{maximize}}
& & \sum_{k = 1}^KR_k \\
& \text{subject to} 
& &  \sum_{k = 1}^{K} p_{k} \leq p_{\max}. 
\end{aligned}
\end{equation}
For the mMIMO scheme the optimization problem becomes
      \begin{equation} \tag{P1-mMIMO}\label{pr:PR-1zf}
\begin{aligned}
& \underset{\vect{p}\succeq 0}{\text{maximize}}
& & \sum_{k = 1}^K\tau  \log_2\left(1 + \left(M-K\right)p_k\beta_{k}\right) \\
& \text{subject to} 
& &  \sum_{k = 1}^{K} p_{k} \leq p_{\max}, 
\end{aligned}
\end{equation}
 which can be solved via the conventional water-filling algorithm described in \cite{telatar1999capacity}.  

For the NOMA scheme the optimization problem instead becomes
\begin{equation} \tag{P1-NOMA}\label{pr:PR-1noma}
\begin{aligned}
& \underset{\vect{p}\succeq 0}{\text{maximize}} 
& & \sum_{k \in \mathcal{K}_c}\tau \log_2\left(1+p_k\beta_k \bar{M}\right)\\   & & & + \sum_{k \in \mathcal{K}_e}\tau   \log_2\left(1+\frac{p_k
	\beta_k }{p_{k-K/2}\beta_k + 1}\right), \\
& \text{subject to} 
& &  \sum_{k = 1}^{K} p_{k} \leq p_{\max}, 
\end{aligned}
\end{equation}
which has a more complicated structure. To solve the problem we first state the following result.
\begin{lemma}\label{lem:NOMAlem1}
	The optimization problem \eqref{pr:PR-1} is maximized if and only if 
	\begin{equation}
	p_k = 0,\quad \forall k \in \mathcal{K}_e,
	\end{equation} 
	under the NOMA scheme. 
\end{lemma}
\begin{IEEEproof}
	See Appendix \ref{ap:ProofLemma1}. 
	\end{IEEEproof}
The importance of Lemma \ref{lem:NOMAlem1} is that to obtain the maximum sum rate, the NOMA scheme does not allocate any power to the cell-edge users, which are effectively dropped from service. Notice that this result holds for any number of antennas and users. Hence, the sum rate maximization problem for NOMA scheme reduces to the following:
      \begin{equation} \tag{P2}\label{pr:PR-2}
\begin{aligned}
& \underset{\vect{p}\succeq 0}{\text{maximize}}
& & \sum_{k = 1}^{K/2}\tau\log_2\left(1 + \bar{M}p_k\beta_k \right) \\
& \text{subject to} 
& &  \sum_{k = 1}^{K/2} p_{k} \leq p_{\max}. 
\end{aligned}
\end{equation}
which, similar to the mMIMO case, can be solved via the conventional water-filling algorithm described in \cite{telatar1999capacity}. 
\begin{remark}
  In a real system, each user would have a rate requirement, albeit it may be low for some users. Considering a setup with rate requirements would result in a lower maximum sum rate for the NOMA scheme. The case where each user has a rate constraint is investigated in Section~\ref{sec:DoWeNeedNOMA}.
\end{remark} 

	\begin{figure}[tb]
	\begin{center}
		\includegraphics[trim=0.5cm 0cm 0cm 0cm,clip=false, scale = .7]{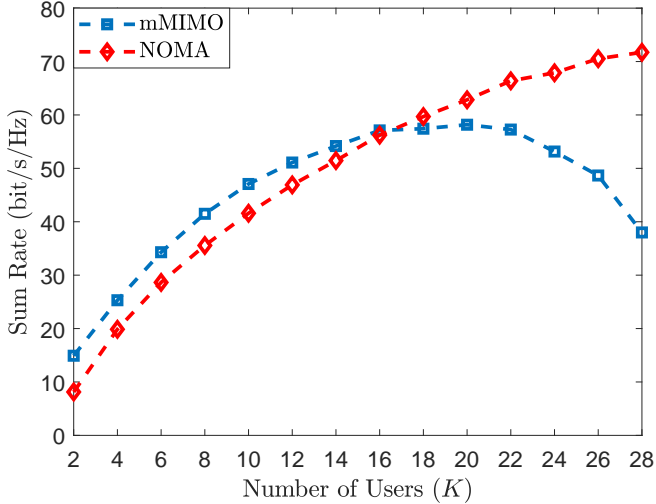}
		\caption{ The average sum rate obtained for the problem defined in \eqref{pr:PR-1} where the average is taken with respect to random user locations. The curves are obtained for various number of users under a setup with $M = 30$ BS antennas and coherence interval length, $T = 100$ symbols. }
		\label{fig:Fig3} 
	\end{center}
\end{figure}

In Fig.~\ref{fig:Fig3}, the average sum rates obtained by mMIMO and NOMA, with respect to number of users, are depicted under a setup with $M = 30$ BS antennas. The cell-center and edge users are uniformly distributed  in certain parts of the cell such that the received SNR of cell-center users are in the range 15--26\,dB whereas for cell-edge users it is from $-5\,$dB to $15\,$dB. The simulation parameters are summarized in Table \ref{tbl:SysParameters}.    
The curves represent the solutions to \eqref{pr:PR-1}. The figure shows that mMIMO provides a higher sum rate than NOMA for $K \leq 16$, while NOMA outperforms mMIMO scheme when $K > 16$. Hence, as $K/M \rightarrow 1$ NOMA becomes superior whereas when $M \gg K$, which is normally what is considered to be massive MIMO in the literature, mMIMO outperforms the NOMA scheme. In other words, NOMA can outperform small-scale multi-user MIMO systems, but not a true mMIMO system.

\begin{remark}
 For the simulations involving optimization problems that lack a closed-form solution, we used CVX, a package for solving convex programs \cite{cvx} to obtain the optimal power allocation vectors.
\end{remark}  

	\begin{figure}[tb]
	\begin{center}
		\includegraphics[trim=0.5cm 0cm 0cm 0cm,clip=false, scale = .7]{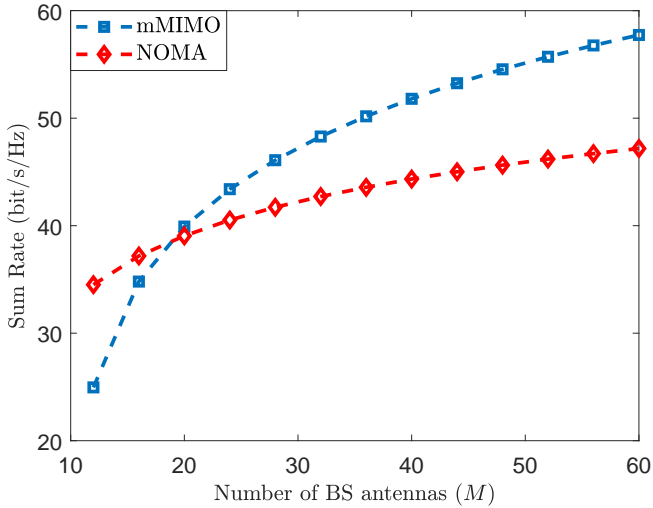}
		\caption{The curves represent the average sum rate obtained for the problem defined in \eqref{pr:PR-1} where the average is taken with respect to random user locations. The average sum rates are obtained for the mMIMO and NOMA schemes as a function of number of BS antennas for $K = 10$ users with a coherence interval length of $T = 100$ symbols.  }
		\label{fig:Fig4} 
	\end{center}
\end{figure}
 
The performance of mMIMO and NOMA with respect to the number of BS antennas are shown in Fig.~\ref{fig:Fig4} for $K=10$. As expected, when $K/M \approx 1$, NOMA provides the better performance, however in the massive MIMO regime, mMIMO is significantly better than the NOMA scheme. Furthermore, even though both approaches benefit from a higher number of BS antennas, the difference between them increases with $M$. 

\subsection{Analytical Example: Two User}

To analytically determine for which values of $M$ and $K$ that mMIMO and NOMA are preferable, we now  investigate \eqref{pr:PR-1} for two users. Even though such an example may seem trivial, the two user case is an important setup to derive basic insights  and it has been investigated in various prior works \cite{xiao2018joint,choi2016power}.
Recall that the sum rate for NOMA is maximized by allocating $p_{\max}$ to the cell-center user, i.e., the optimal power vector that maximizes sum rate for NOMA is $\vect{p}^* = [p_{\max},\,0]^T$ which is shown by Lemma \ref{lem:NOMAlem1}. For mMIMO, we have the following result. 
\begin{lemma}\label{lem:ZF_2user}
	Consider the optimization problem \eqref{pr:PR-1} for two users with the mMIMO scheme. Then, the optimum power transmit powers are
	\begin{eqnarray}
	p_1 &=& \min\left(p_{\max},~\frac{\beta_1 - \beta_2 + p_{\max}\beta_1\beta_2\left(M - 2\right)}{2\beta_1\beta_2\left(M-2\right)}\right), \label{eq:ZFoptP1}\\
	p_2 &=& p_{\max} - p_1. \label{eq:ZFoptP2}
	\end{eqnarray}  	
\end{lemma}
\begin{IEEEproof}
	See Appendix \ref{ap:ProofLemmaZF_2user}.
\end{IEEEproof}	

Using \eqref{eq:ZFoptP1} and \eqref{eq:ZFoptP2}, the maximum sum rate with mMIMO is 
\begin{equation} \label{eq:ZFmax_2user}
R_{\max}^{\mathrm{mMIMO}} = \tau \log_2\left(\frac{\left(\beta_1 + \beta_2 + p_{\max}\beta_1\beta_2\left(M -2\right)\right)^2}{4\beta_1\beta_2}\right),
\end{equation}
assuming that  	
\begin{equation} \label{eq:ZFcond}
p_{\max} \geq \frac{\beta_1- \beta_2}{\beta_1\beta_2\left(M-2\right)},
\end{equation}
holds. If \eqref{eq:ZFcond} is not satisfied, we have 
\begin{equation} \label{eq:ZFmax_2user2}
R_{\max}^{\mathrm{mMIMO}} = \tau \log_2\left( 1 + p_{\max}\beta_1\left(M - 2\right)\right).
\end{equation}
Note that the condition given in \eqref{eq:ZFcond} determines whether cell-edge user is dropped or not, i.e., if \eqref{eq:ZFcond} holds, $p_2 > 0$. 

For NOMA we have 
\begin{equation} \label{eq:NOMAmax_2user}
R_{\max}^{\mathrm{NOMA}} = \tau\log_2\left(1+ p_{\max}\beta_1 M\right).
\end{equation}
In the NOMA scheme, the sum rate scales with $M$ while with mMIMO it scales with $\left(M - 2\right)^2$ assuming that \eqref{eq:ZFcond} holds. This suggest that when $M$ is small, NOMA may provide a better performance, however mMIMO will always outperform NOMA scheme when $M$ becomes sufficiently large. For the two user case, we can compare these two schemes and find an expression for the number of BS antennas where mMIMO starts to provide a higher sum rate.

\begin{lemma}\label{lem:CrossoverM}
	Consider the optimization problem \eqref{pr:PR-1} for two users and assume that \eqref{eq:ZFcond} is satisfied
	to ensure that $p_2 > 0$ with mMIMO.
	Let $M^* \in \mathbb{Z}^+$ denote the minimum number of BS antennas such that 
	\begin{equation}
	R_{\max}^{\mathrm{mMIMO}} \geq R_{\max}^{\mathrm{NOMA}}, \quad \forall M \geq M^*.
	\end{equation}
	Then, 
	\begin{equation} \label{eq:Mstar}
	M^* = \lceil{\tilde{M}\rceil},
    \end{equation}
    where 
    \begin{equation}
     \tilde{M}= 	2 + \frac{\beta_1 - \beta_2}{p_{\max}\beta_1\beta_2} + \frac{2\sqrt{2}}{\sqrt{p_{\max}\beta_2}},      
    \end{equation}
    and $\lceil{\cdot\rceil}$ is the ceiling function.
\end{lemma}
\begin{IEEEproof}
	See Appendix \ref{ap:prCrossover}.
	\end{IEEEproof}

\begin{figure}[tb]
	\begin{center}
		\includegraphics[trim=0.5cm 0cm 0cm 0cm,clip=false, scale = .7]{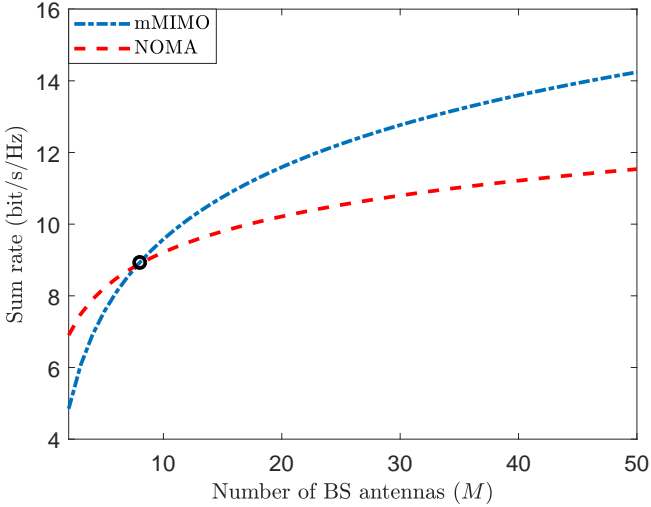}
		\caption{ Maximum sum rate as a function of number of BS antennas. Black circle shows $M^*$ defined in \eqref{eq:Mstar}. }
		\label{fig:Fig2} 
	\end{center}
\end{figure}

Lemma \ref{lem:CrossoverM} provides insights into the relative performance of the mMIMO and NOMA schemes. First, notice that NOMA benefits from the difference between large-scale fading coefficients. Second, a smaller $\beta_2$ also favors the NOMA scheme which is in alignment with the grouping of cell-edge and center users. Finally, high SNR favors the mMIMO scheme. Fig.~\ref{fig:Fig2} illustrates the sum rates obtained for two user case as well as $M^*$ given in \eqref{eq:Mstar}, where mMIMO becomes superior for $M\geq 9$, which is much smaller than what is normally referred to as massive MIMO. For this particular example, cell-center user is located at a distance of $100\,$m to the BS, whereas the cell-edge user is located at the edge of the cell.

\section{Performance Analysis for LOS} \label{sec:LOS}

In this section, we investigate the performance of the NOMA and mMIMO schemes in a LOS setup. Contrary to the NLOS case, the channel between the transmitter and the receiver deterministic and perfectly known at the BS.

The LOS channel between the BS and the user $k$ is 
\begin{equation}\label{eq:LOSchannel}
\vect{g}_k = \sqrt{\beta_k}\vect{h}_k, \quad \forall k = 1, \ldots,K, 
\end{equation}
where in case of a uniform linear array (ULA) we have
\begin{equation} \label{eq:LOSarray}
\vect{h}_k = \left[1\,\, e^{j 2\pi \frac{d}{\lambda}\sin{\phi_k}}, \ldots, e^{j 2\pi (M-1)\frac{d}{\lambda}\sin{\phi_k}} \right]^T, \, \forall k = 1, \ldots,K.
\end{equation}
Here, $d$ represents the distance between two adjacent antennas in the ULA; $\lambda$ is the carrier wavelength; $\phi_k$ is the angle of departure from the BS station to the $k$th user, relative to the array boresight.\footnote{In \eqref{eq:LOSarray}, a term to model the phase rotation of all antennas in the LOS channel model, can also be included. However, this term does not affect the achievable rates and is thus omitted.} It is important to note that we consider a uniform linear array for the LOS case and \eqref{eq:LOSarray} is valid under the assumption that users are located in the far field of the ULA and there is no scattering.

\begin{remark}
The LOS channel defined in \eqref{eq:LOSarray} is only one of the many ways to generate $\vect{h}_k$ and is valid for ULAs only. This is the model that we utilize in the simulations. Based on the array geometry, $\vect{h}_k$ may be generated in different ways \cite[Section~7.3]{massivemimobook}. However, the rate expressions provided are valid for any other array setups. 
\end{remark}

\subsection{Achievable Rates}

Similar to the NLOS case, the received signal at user $k$ is given by \eqref{eq:RecSignalGeneral}. We once again consider the NOMA and mMIMO  schemes.
For the mMIMO scheme, we have a beamforming matrix
 $\vect{V} = [\vect{v}_1, \ldots,\vect{v}_K]$ given by
\begin{equation}\label{def:ZFprecoder}
\vect{V} = \vect{H}\left(\vect{H}^H\vect{H}\right)^{-1}, 
\end{equation}
where $\vect{H} = [\vect{h}_1, \ldots, \vect{h}_K]$ is the $M \times K$ channel matrix. The achievable rate of user $k$ is 
\begin{equation}
R_k = \log_2\left(1 + \text{SINR}_k\right)
\end{equation}
where the SINR of user $k$ when treating interference as noise, is 
\begin{equation} \label{def:SINR_LOS_MIMO}
\text{SINR}_k = 
\frac{p_k\beta_k}{ \left[\left(\vect{H}^H\vect{H}\right)^{-1}\right]_{k',k'} },\quad \text{for mMIMO}.
\end{equation}
Here, $[\vect{A}]_{k,k}$ denotes the $k$th diagonal element of matrix $\vect{A}$. 

For the NOMA scheme, the achievable rate for cell-center user $k$ is given by 
\begin{flalign}
&R_{k}^{\mathrm{NOMA}}= \log_2\left(1+\frac{p_k
	\beta_k |\vect{h}_k^T\vect{v}_k|^2}{\beta_k\hspace{-3mm}\sum\limits_{\substack{k'\neq k,\\ k' \neq k+K/2}}^K \hspace{-3mm} p_{k'}|\vect{h}_k^T\vect{v}_{k'}|^2+ 1}\right), \forall k \in \mathcal{K}_c,\label{eq:rateNOMAccLOS}
\end{flalign}
and 
\begin{equation}\label{eq:rateNOMAceLOS}
R_{k}^{\mathrm{NOMA}}=\log_2 \left(1 + \frac{p_{k+K/2}\beta_{k}|\vect{h}_k^T \vect{v}_{k + K/2}|^2}{\beta_k \sum_{k' \neq k + K/2}^{K} p_{k'}|\vect{h}_k^T \vect{v}_{k'}|^2 + 1} \right), \forall k \in \mathcal{K}_e,
\end{equation} 
where we assumed that the cell-center user first decodes the data symbols of cell-edge user and perform SIC similar to the NLOS case. The interference from other groups are treated as noise.

\subsection{Performance Comparison}

In this section, we compare the performance of the NOMA and mMIMO schemes in the LOS setup. Similar to the NLOS case, we consider the maximum sum rate problem defined in \eqref{pr:PR-1}. First, we focus on the NOMA scheme and state the following result.
\begin{lemma}
	The sum rate optimization problem defined in \eqref{pr:PR-1} is maximized if and only if 
	\begin{equation}
	p_k = 0,\quad \forall k \in \mathcal{K}_e,
	\end{equation} 
	with the NOMA scheme under the LOS setup. 
\end{lemma}
\begin{IEEEproof}
The proof is similar to the NLOS case, the only difference is the definition of $\vect{h}_k$, and therefore is omitted. 
\end{IEEEproof}

The conclusion is once again that NOMA scheme should allocate all its power to the cell-center users in order to maximize the sum rate.
For the mMIMO scheme, the solution to \eqref{pr:PR-1} can be obtained via water-filling methods. 

	\begin{figure}[tb]
	\begin{center}
			\includegraphics[trim=0.5cm 0cm 0cm 0cm,clip=false, scale = .7]{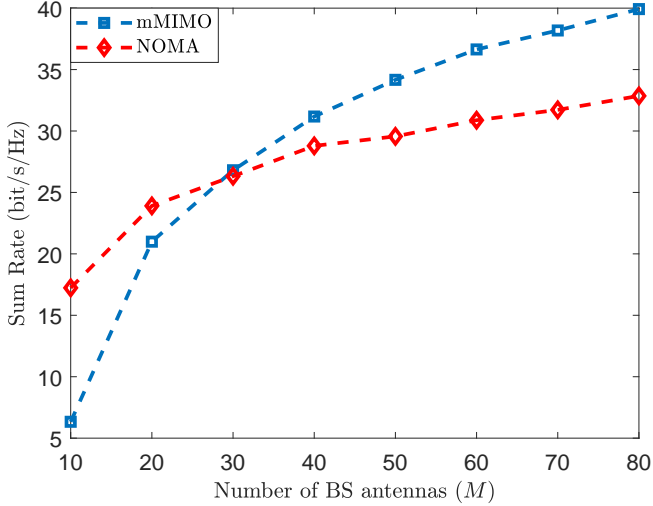}
		\caption{ Comparison of the average of the maximum sum rate solutions to the optimization problem defined in \eqref{pr:PR-1} as a function of number of BS antennas under LOS setup for $K = 10$ users.}
		\label{fig:Fig5} 
	\end{center}
\end{figure}

The performance of mMIMO and NOMA schemes under LOS setup with $K = 10$ users, are depicted in Fig.~\ref{fig:Fig5}. The user angles are generated randomly, i.e., $\phi_k\sim\mathsf{U}(0,2\pi)$ for $k \in \mathcal{K}$. Similar to the NLOS case, the users are uniformly distributed. The details are provided in Table \ref{tbl:SysParameters}. Higher number of BS antennas favors the mMIMO schemes more compared to the NOMA scheme. When $M/K \approx 1$, the NOMA scheme provides the best performance. However, as $M$ increases, mMIMO outperforms NOMA. This is consistent with the observations made in the NLOS case.

	\begin{figure}[tb]
	\begin{center}
		\includegraphics[trim=0.5cm 0cm 0cm 0cm,clip=false, scale = .7]{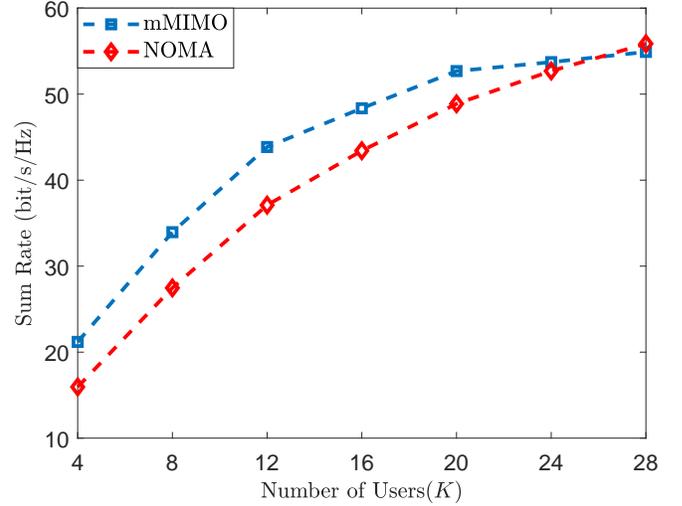}
		\caption{ Comparison of the average of the maximum sum rate solutions to the optimization problem defined in \eqref{pr:PR-1} with respect to the number of users under LOS setup for $M = 75$ BS antennas.}
		\label{fig:Fig6} 
	\end{center}
\end{figure}  

Finally, Fig.~\ref{fig:Fig6} illustrates the sum rate as a function of number of users for $M=75$. Similar to the previous cases, mMIMO outperforms the NOMA scheme for most values of $K$, but NOMA becomes the better choice for very large $K$ values (in this case $K > 28$).

\section{Do We Need NOMA in Massive MIMO Systems?}\label{sec:DoWeNeedNOMA}

Recall that massive MIMO usually refers to scenarios with $M \gg K$ (i.e., many more antennas than users). 
The results presented in Sections \ref{sec:LOS} and \ref{sec:NLOS} indicate that the NOMA scheme brings benefits over the mMIMO scheme when $M/K \approx 1$, but not in the typical massive MIMO scenarios. However, this is not a conclusive observation. We will now demonstrate that there are cases where the NOMA scheme can provide gains also in massive MIMO scenarios, and we will then demonstrate how to utilize that in a hybrid mMIMO-NOMA scheme.

	\begin{figure}[tb]
	\begin{center}\vspace{-5mm}
		\includegraphics[trim=0.9cm 0cm 0cm 0cm,clip=false, scale = .65]{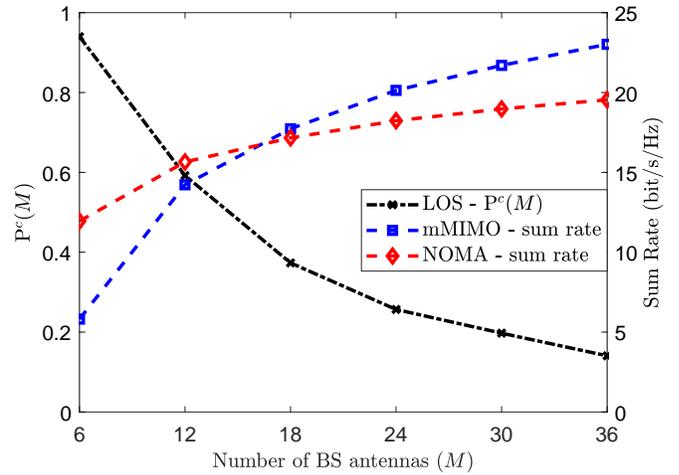}
		\caption{On the left axis, $\text{P}^c(M)$ is depicted as a function of number of BS antennas under LOS setup, with $K = 6$ users. On the right, the average sum rate obtained for the mMIMO and NOMA schemes are shown. }
		\label{fig:Fig7} 
	\end{center}
\end{figure} 

Let $\text{P}^c(M)$ denote the probability that a random realization of the user locations lead to a sum rate that is higher with the NOMA scheme than with the mMIMO scheme, for a given $M$. In Fig.~\ref{fig:Fig7}, $\text{P}^c(M)$ is depicted for $K = 6$ users in the LOS setup. On the right axis, the maximum sum rates for the LOS case is shown for the mMIMO and NOMA schemes. Even though, a higher sum rate is achieved with mMIMO on the average when $M > 18$, there is still a significant probability that NOMA outperforms mMIMO in a particular realization of the user locations. The use of NOMA in those cases would result in a higher average sum rate. However, this would require a system with the flexibility to switch between NOMA and mMIMO depending on which users the BS is serving for the moment.

\subsection{In Which Situations Do We Need NOMA?}
  
The favorable propagation concept in the massive MIMO literature says that the users' channels become mutually orthogonal when $M \to \infty$ \cite{ngo2014aspects,massivemimobook}. However, since $M$ is finite in any practical system, the probability that two users have similar channels is non-negligible. Users with similar channel conditions may deteriorate the performance of mMIMO schemes. With this in mind, there are several prior works that proposed dropping users that have similar channels to other users, in an effort to obtain a user set where all the channels are approximately mutually orthogonal \cite{ngo2014aspects},\cite{yang2017massive}. 

On the other hand, in the NOMA scheme, two users in the same group ideally should have identical channels, so that one beamforming vector fits both users, and different large-scale fading coefficients. Hence, in the massive MIMO regime with a finite $M$ that satisfies $M\gg K$, we can benefit from employing NOMA when two users have nearly parallel channels---this can happen in both LOS and NLOS scenarios, but is more probable in the former situation. 

\subsection{How to Use NOMA in Massive MIMO?}

In this section, we propose a hybrid approach, referred to as hybrid mMIMO-NOMA (HmNOMA), which combines the advantages of the two schemes, mMIMO and NOMA, as follows. After the channel vectors have been estimated, based on the similarity of these channels, the BS either creates groups where each group utilizes the NOMA scheme or creates beamforming vectors using the conventional mMIMO scheme with ZF beamforming. Note that the performance of the NOMA scheme increases with the correlation of the channels of the users in a group, since the beamformer will then fit both user. The correlation between the channels of users $i$ and $j$ is 
\begin{equation} \label{eq:corr}
\rho_{i,j} = \frac{|\vect{h}_i^H\vect{h}_j|}{\|\vect{h}_i\|\|\vect{h}_j\|}.
\end{equation}
For the LOS case, \eqref{eq:corr} reduces to
\begin{equation}
\rho_{i,j} = \frac{1}{M}\left|\frac{1 - e^{j2\pi\frac{d}{\lambda} \left(\sin(\phi_i) - \sin(\phi_j)\right)M}}{1 - e^{j2\pi\frac{d}{\lambda} \left(\sin(\phi_i) - \sin(\phi_j)\right)}}\right|
\end{equation} which attains its maximum as $|\sin(\phi_i) - \sin(\phi_j)| \rightarrow 0$. Hence, we utilize 
\begin{equation}
d_{i,j}^{\text{LOS}} = |\sin(\phi_i) - \sin(\phi_j)|,
\end{equation} as a distance measure in the LOS case, whereas in the NLOS case, we have
\begin{equation}
d_{i,j}^{\text{NLOS}} = \rho_{i,j}.
\end{equation}
In order to employ the hybrid approach, the pairing algorithm must be carried out at each coherence interval based on the channel estimates. Moreover, for the NOMA groups downlink pilots must be transmitted. In the subsequent numerical analysis, we assume that downlink pilots are transmitted without any cost and the perfect CSI is available at the users which employs NOMA.

In the LOS case, user $i \in \mathcal{K}_c$ and user $j \in \mathcal{K}_e$ are grouped together if
  \begin{equation}\label{eq:LOSdistMeasure}
  |\sin(\phi_i) - \sin(\phi_j)| \leq \nu,
  \end{equation}
  where the threshold $\nu$ is a design variable.
A simple pairing algorithm for the LOS case is described in Algorithm \ref{alg:Alg1LOS}. For the NLOS case, $d_{i,j}^{\text{NLOS}}$ is computed based on the correlation of the channels and the pairing is accomplished by grouping users with maximum correlation.
 This algorithm is similar to the one given in \cite{chen2016application}, with the difference that the pairing decision is based on a predefined threshold, $\nu$ and the distance measure defined in \eqref{eq:LOSdistMeasure}.   

\begin{algorithm}
	\caption{Pairing Algorithm - LOS}\label{alg:Alg1LOS}
	\begin{algorithmic}[1]
	 \Statex \textbf{INPUT}: $\phi_1, \ldots, \phi_K$ 
	 \Statex \textbf{OUTPUT}: $\Pi$ $\gets \{\}$, \hfill // set of pairs 
	 \State \quad\textbf{for} i = $1:$$K/2$ \hfill // for all cell-center users
	 \State \quad\quad \textbf{for} j = $K/2 + 1: K$ \hfill // for all cell-edge users
	 \State \quad\quad\quad \text{\textbf{if}} user $j$ is not paired
	 \State \quad\quad\quad \quad compute $d_{i,j}^{\text{LOS}}$ 
	 \State \quad\quad\quad \text{\textbf{end if}}
	 \State \quad\quad \textbf{end for}
	 \State \quad$ j^* =\argmin\limits_j d_{i,j}$  \hfill //find the most similar channel 
	 \State \quad\quad\quad \textbf{if} $d_{i,j^*} < \nu$ \hfill // Compare with a threshold 
	 \State \quad\quad\quad\quad $\Pi$ = $\Pi \cup (i,j^*)$ 
	 \State \quad\quad\quad \textbf{end if}
	 \State \quad\textbf{end for}
	\end{algorithmic}
\end{algorithm}

\begin{remark}
 Note that, there is no claim of optimality of this simple pairing algorithm. There are various additional factors that may be considered while pairing users, such as the large-scale coefficient difference, rate requirements, \etc. We leave the optimal user pairing for NOMA in massive MIMO setups as future work.      
\end{remark}
    
\begin{figure}[tb]
    	\begin{center}
    		\includegraphics[trim=0.5cm 0cm 0cm 0cm,clip=false, scale = .7]{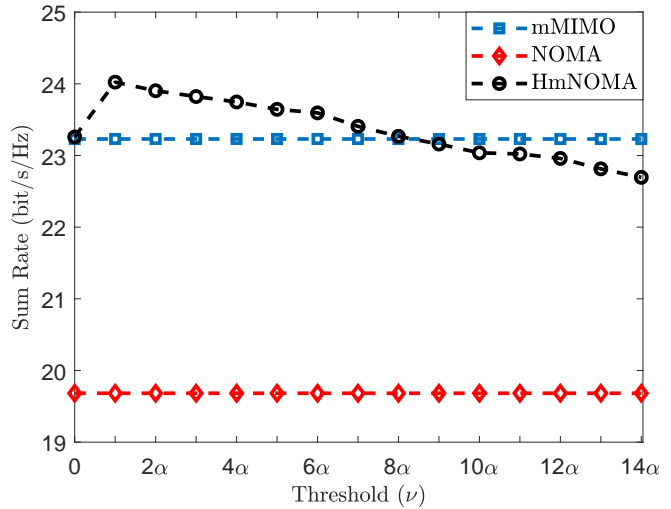}
    		\caption{The hybrid NOMA (HmNOMA) approach compared with the mMIMO and NOMA schemes in terms of the average sum rate for various threshold values under a setup with $K = 6$ users and $M = 36$ BS antennas. Here, $\alpha = 1/(2M)$ and only HmNOMA curve is a function of the threshold $\nu$ as other approaches do not utilize a threshold.  } 
    		\label{fig:Fig8} 
    	\end{center}
\end{figure} 
 
In Fig.~\ref{fig:Fig8}, the sum rate obtained with the HmNOMA scheme is compared with the mMIMO and NOMA schemes in a setup with $K = 6$ users and $M = 12$ BS antennas. The curves are obtained for various values of the threshold $\nu$ for the HmNOMA, whereas the mMIMO and NOMA schemes do not utilize any threshold and therefore have straight lines. Here, $\alpha = 1/2M$ is chosen as it has been shown in \cite{ngo2014aspects} that if there exist any pair of users where $|\sin(\phi_i) - \sin(\phi_j)|$ is in the order of $1/M$, the favorable propagation assumption does not hold. As expected, mMIMO performs better that NOMA in terms of average sum rate. However, HmNOMA performs the best when $\nu \leq 10\alpha$. The HmNOMA scheme attains its maximum sum rate when the threshold is at $\nu = 1/2M$, which suggest that for users with very small angle differences, the use of NOMA is recommended.      

For the sum rate maximization problem, the HmNOMA scheme boils down to dropping the cell-edge users that has similar channels to the cell-center users, as we have observed earlier for the pure NOMA scheme. However, in a real system each user has a rate constraint that must be satisfied. To assess the performance of different approaches with rate constraints, we consider the following problem: 
 	\begin{equation} \tag{P3}\label{pr:PR-3}
 	\begin{aligned}
 	& \underset{\vect{p}\succeq 0}{\text{maximize}}
 	& & \mu \\
 	& \text{subject to} 
 	& & R_k \geq \mu \quad\quad~ \forall k \in \mathcal{K}_c,  \\ 
 	& & & R_j \geq c\mu \quad\quad \forall j \in \mathcal{K}_e,  \\ 
 	& & & \sum_{k = 1}^{K} p_{k} \leq p_{\max}. 
 	\end{aligned}
 	\end{equation}
In \eqref{pr:PR-3}, the goal is to provide each cell-center user with a target rate $\mu$ and each cell-center user with a smaller target rate $c\mu$ where $0<c <1$. This ensures that the cell-edge users are served with a smaller rate compared to the cell-center users instead of being dropped.

\begin{figure}[tb]
	\begin{center}
		\includegraphics[trim=0.2cm 0cm 0cm 0cm,clip=true, scale = .7]{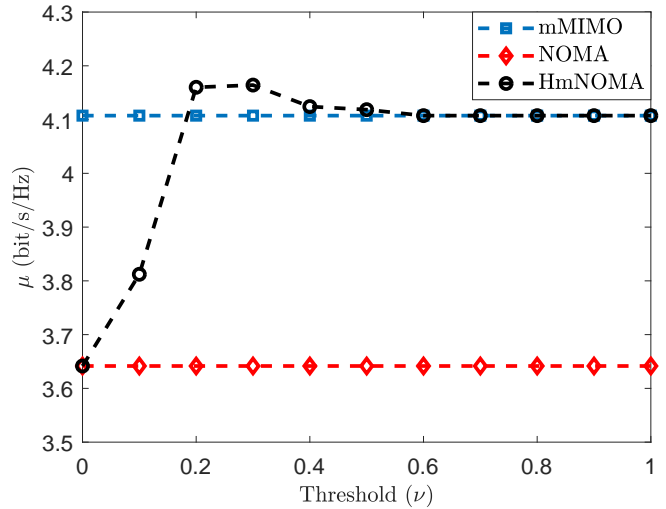}
		\caption{The average of the optimal solutions obtained for \eqref{pr:PR-3} via the mMIMO, HmNOMA, and NOMA schemes. The curves are obtained under a NLOS setup with $K = 10$ users and $M = 12$ BS antennas. There is constraint to ensure that cell-center users achieve $100$ times more SINR than the cell-edge users. }
		\label{fig:Fig9} 
	\end{center}
\end{figure} 

Fig.~\ref{fig:Fig9} illustrates the performance of the mMIMO, NOMA, and HmNOMA schemes for \eqref{pr:PR-3} under a NLOS setup. In this example, there are $K = 10$ users and $M = 12$ antennas. An SINR constraint is imposed such that the cell-center users achieve $100$ times more SINR than the cell-edge users which results in a rate constraint with $c \approx 0.05$ in \eqref{pr:PR-3}. For the HmNOMA, we employ a modified pairing algorithm defined in Algorithm \eqref{alg:Alg1LOS}, which is suitable for the NLOS case.  HmNOMA outperforms mMIMO when the threshold is $0.2 < \nu < 0.5$, which shows that NOMA can provide gains when the users channels are similar, yet the gains are around $1.4\%$. For the higher threshold values, the probability of having two users with such a high correlation is very low and therefore, HmNOMA performs similar to mMIMO at higher threshold values. 

\begin{figure}[tb]
	\begin{center} \vspace{-0.5cm}
		\includegraphics[trim=0.2cm 0cm 0cm 0cm,clip=false, scale = .7] {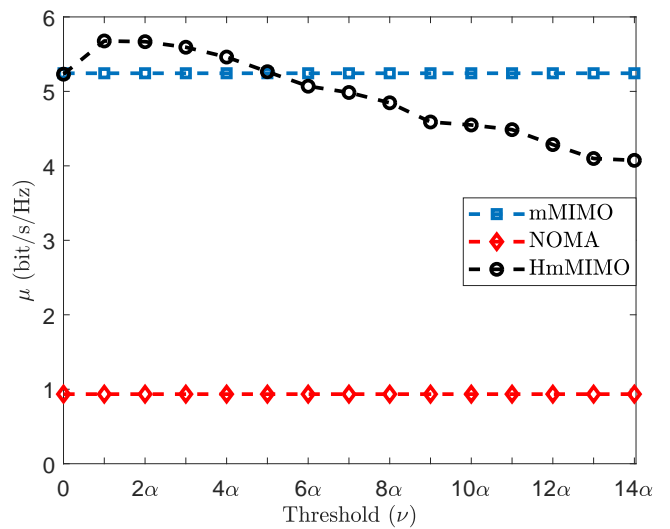}
		\caption{The average of the optimal solutions obtained for \eqref{pr:PR-3} via the mMIMO, HmNOMA, and NOMA schemes. The curves are obtained under a setup with $K = 6$ users, $M = 36$ BS antennas and, $\alpha = 1/(2M)$.  } \vspace{-.5cm}
		\label{fig:Fig10} 
	\end{center}
\end{figure} 

As a final example, we consider \eqref{pr:PR-3} under a LOS setup. For this particular example, there are $K= 6$ users and $M = 36$ antennas. Similar to the previous example, an SINR constraint is imposed such that the cell-center users achieve $100$ times more SINR than cell-edge users. Although mMIMO heavily outperforms NOMA, HmNOMA still provides the best performance when the threshold is $\alpha < \nu < 5\alpha$. We see that HmNOMA provides gains more than $8\%$ for this particular LOS example.

\section{Conclusion}

NOMA is known to provide considerably higher rates than OMA, but how does it perform compared with other popular multiple access schemes in 5G scenarios?
In this work, we have investigated the gains that NOMA can provide in massive MIMO setups, with a large number of antennas compared to the number of users, and compared it with a standard approach from the massive MIMO literature. 
More precisely, instead of the conventional comparison between NOMA and OMA, we compare two non-orthogonal approaches: power-domain NOMA and mMIMO using ZF. 
This comparison is  practically important since BSs equipped with many antennas are already being deployed in LTE networks and will be the norm when the 5G deployments take off. It is important to remember that the conventional mMIMO techniques, which use spatial multiplexing to serve multiple users on the same time/frequency/code resources and rely on beamforming to partially suppress inter-user interference, cannot be considered OMA schemes.

In the first part of the paper, the performance of two standard NOMA and mMIMO schemes are compared in a NLOS setup. We prove analytically that when $M\gg K$, mMIMO achieves the highest average sum rate. However, in cases where $M \approx K$, the NOMA scheme is better. A closed-form expression for the minimum number of BS antennas such that the mMIMO scheme outperforms NOMA, is derived in the two user case, which shows that NOMA benefits from a smaller large-scale fading coefficient for the cell-edge user and a large difference between the large-scale fading coefficients. Similar observations were made in the LOS case.

In the second part of the paper, we demonstrate that even though the mMIMO scheme significantly outperforms NOMA in terms of average sum rate when $M\gg K$, NOMA can still be useful in massive MIMO systems. 
By analyzing the probability that the NOMA scheme outperforms the mMIMO scheme for a random set of user locations, we observed that there is a non-negligible probability for this to happen in LOS cases.
Inspired by this result, we provide a hybrid approach, HmNOMA, along with a  pairing algorithm to form user groups based on the similarity of the users' channels. The HmNOMA identifies those users that would benefit from power-domain NOMA operation, while the rest are served as in standard mMIMO. This hybrid approach outperforms the standalone  NOMA and mMIMO schemes. However, in the NLOS case, HmNOMA provides only modest gain compared to the standalone mMIMO scheme and those gains should be taken with caution as we have made several assumptions when developing the rate expressions that favors NOMA.

This work considers the two extremes in terms of channel models: the NLOS case with i.i.d. Rayleigh fading and the LOS case with deterministic channels. A combination of the NOMA and mMIMO schemes seems to provide gains in both cases, albeit the gains are substantially higher in the LOS case. In a real system, it is likely to have a channel which is a combination of these two cases; for example, a LOS scenario that also contains fading or a NLOS case where the small-scale fading exhibits strong spatial correlation. Hence, it can be concluded that a mMIMO system that supports NOMA and has the ability to switch between the NOMA and mMIMO schemes can perform better than a system that only employs one of the schemes.   

\appendices

\section{Proof of Lemma \ref{lem:NOMAlem1}} \label{ap:ProofLemma1}
Consider the group $k$, for which we have 
\begin{equation}
p_{k,1} + p_{k,2} = P
\end{equation}
for some positive transmit power $P$, denoting the total transmission power assigned to this group. Here, $p_{k,1}$ and $p_{k,2}$ denote the transmit power of the cell-center and the cell-edge user, respectively. The sum rate is given by 
\begin{equation}
f\left(p_{k,1},p_{k,2}\right) = \log_2\left(1 + \bar{M}\beta_1p_{k,1}\right) + \log_2 \left(1 + \frac{p_{k,2}\beta_2}{1 + p_{k,1}\beta_2}\right)
\end{equation}
and the derivative with respect to $p_{k,1}$ is 
\begin{eqnarray}
\frac{df\left(p_{k,1}, P - p_{k,1}\right)}{dp_{k,1}}\hspace{-3mm} &=& \hspace{-3mm}\frac{1}{\ln(2)}\left( \frac{\bar{M}\beta_1}{1 + \bar{M}\beta_1p_{k,1}} - \frac{\beta_2}{1 + \beta_2p_{k,1}}\right), \nonumber \\
&=& \hspace{-2mm}\frac{\bar{M}\beta_1 - \beta_2}{\ln(2)\left(1 + \bar{M}\beta_1p_{k,1} \right)\left(1 + \beta_2p_{k,1}\right)},
\end{eqnarray}
which is always positive if $\bar{M}\beta_1 > \beta_2$. Since, $\bar{M} \geq 1$ and $\beta_1 > \beta_2$ by definition, we have
\begin{equation} 
\underset{p_1,p_2\succeq 0}{\text{max}}f\left(p_{k,1},p_{k,2}\right) = f\left(P, 0\right), 
\end{equation}   
which concludes the proof since the result is independent of the actual value of $P$.
\section{Proof of Lemma \ref{lem:ZF_2user}} \label{ap:ProofLemmaZF_2user}
The sum rate for two users with ZF is given by 
\begin{equation}
g\left(p_{1},p_{2}\right) = \log_2\left(1 + p_1\beta_1\bar{M}\right) + \log_2\left(1 + p_2\beta_2\bar{M}\right), 
\end{equation}
where $\bar{M} = M - 2$ and the derivative with respect to $p_1$ is 
\begin{equation}
\frac{dg\left(p_{1}, p_{\max} - p_{1}\right)}{dp_{1}}  =\frac{1}{\ln(2)} \left(\frac{\bar{M}\beta_1}{1 + \bar{M}\beta_1p_1} - \frac{\beta_2}{1 + \beta_2p_1}\right). 
\end{equation}
The second derivative of $g\left(p_{1}, p_{\max} - p_{1}\right)$ with respect to $p_1$ is negative. Hence, the optimum power allocation can be found by equating the derivative to zero, leading to:
\begin{equation}
\bar{M}\left(\beta_1 - 
\beta_2\right) + \left(p_{\max} - 2p_1\right)\beta_1\beta_2\bar{M}^2 = 0,
\end{equation}
which can be solved for $p_1$ to obtain
\begin{equation}
p_1 = \frac{\beta_1 - \beta_2 + p_{\max}\beta_1\beta_2\left(M - 2\right)}{2\beta_1\beta_2\left(M-2\right)}.
\end{equation} 
Since, $g\left(p_{1}, p_{\max} - p_{1}\right)$ is a concave function, the optimal transmit power for the cell-center user with a power constraint, is 
\begin{equation}
p_1 = \min\left(p_{\max},~\frac{\beta_1 - \beta_2 + p_{\max}\beta_1\beta_2\left(M - 2\right)}{2\beta_1\beta_2\left(M-2\right)}\right),
\end{equation} 
which concludes the proof. 

\begin{table}
	\centering
	\caption{Simulation Parameters}
	\label{tbl:SysParameters}
	\begin{tabular}{l|l}
		\hline
		\textbf{System Parameter} & \textbf{Value} \\ \hline
		Path and penetration loss at distance $d$ (km) & 130 + 37.6 $\log_{10}(d)$ \\ 
		Cell edge length           & $350$ m          \\
		Center user minimum distance      & $50$ m           \\ 
		Center user maximum distance      & $100$ m \\
		Edge user minimum distance      & $100$ m           \\ 
		Edge user maximum distance      & $350$ m \\		
		Coherence interval length (in symbols) & $100$ \\
		Received SNR at cell-center users & [$15$, $26$] dB \\
		Received SNR at cell-edge users & [$-5$, $15$] dB \\
		\hline 
	\end{tabular}
\end{table}

\section{Proof of Lemma \ref{lem:CrossoverM}} \label{ap:prCrossover}
The maximum sum rates for the mMIMO and NOMA schemes in the two user setup are given by \eqref{eq:ZFmax_2user} and \eqref{eq:NOMAmax_2user}. In order to find $M^*$,														 we compare the sum rates as follows:
\begin{eqnarray}
\tau \log_2\left(\frac{\left(\beta_1 + \beta_2 + P\beta_1\beta_2\left(M -2\right)\right)^2}{4\beta_1\beta_2}\right)\hspace{-3mm} &\underset{\mathrm{NOMA}}{\overset{\mathrm{mMIMO}}{\gtrless}}&\hspace{-4.5mm}  \tau\log_2\left(1+ P\beta_1 M\right), \nonumber \\
\frac{\left(\beta_1 + \beta_2 + P\beta_1\beta_2\left(M -2\right)\right)^2}{4\beta_1\beta_2}\hspace{-2mm} &\underset{\mathrm{NOMA}}{\overset{\mathrm{mMIMO}}{\gtrless}}&\hspace{-3mm}  1+ P\beta_1 M, \label{eq:ZFvsNOMApr1} 
\end{eqnarray}
where $P = p_{\max}$. \eqref{eq:ZFvsNOMApr1} can be rearranged to obtain
\begin{eqnarray}
\left(M - 2 - \frac{\beta_1 - \beta_2}{P\beta_1\beta_2}\right)^2 - \frac{8}{P\beta_2} &\underset{\mathrm{NOMA}}{\overset{\mathrm{mMIMO}}{\gtrless}}& 0 \\
\left(M - a_1\right)\left(M - a_2\right) &\underset{\mathrm{NOMA}}{\overset{\mathrm{mMIMO}}{\gtrless}}& 0
\end{eqnarray}
where 
\begin{eqnarray}
a_1 &=& 2 + \frac{\left(\beta_1- \beta_2\right)}{P\beta_1\beta_2} - \frac{2\sqrt{2}}{\sqrt{P\beta_2}},\\
a_2 &=&  2 + \frac{\left(\beta_1- \beta_2\right)}{P\beta_1\beta_2} + \frac{2\sqrt{2}}{\sqrt{P\beta_2}}.
\end{eqnarray}
Since, $a_1 < a_2$, for any $M > a_2$, the mMIMO scheme provides a higher sum rate than the NOMA scheme. 


\bibliographystyle{IEEEtran}
\bibliography{IEEEabrv,References}

\end{document}